\documentclass[aps,showpacs,nofootinbib]{revtex4}
\usepackage{amsmath}
\usepackage{amsfonts}
\usepackage{amssymb}
\usepackage{color}
\usepackage{mathrsfs}

\def\bc{\begin{center}}
\def\nno{\nonumber}
\def\ec{\end{center}}
\def\be{\begin{eqnarray}}
\def\ee{\end{eqnarray}}

\newcommand{\omits}[1]{}

\definecolor{dyellow}{rgb}{1.,0.8,.0}
\definecolor{myblue}{rgb}{.1,.1,.7}
\definecolor{dcyan}{rgb}{.0,.6,.6}
\definecolor{dmagenta}{rgb}{0.6,0.0,0.6}
\definecolor{brown}{rgb}{0.6,0.2,0.}
\definecolor{darkblue}{rgb}{.0,.0,0.5}
\definecolor{darkred}{rgb}{0.75,0.0,0.0}
\definecolor{orange}{rgb}{1.,.6,.0}
\definecolor{dorange}{rgb}{0.8,.4,.0}
\definecolor{lightgray}{rgb}{0.7,0.7,0.7}
\definecolor{darkgreen}{rgb}{0.0,0.6,0.0}
\definecolor{purple}{rgb}{.4,.0,.4}
\def\Dl{\Delta}
\def\La{\Lambda}

\def\dl{\delta}
\def\eps{\epsilon}

\def\si{\sigma}

\def\d#1#2{\frac{\displaystyle #1}{\displaystyle #2}}

\newcommand{\dS}{$d{\cal S}$}
\newcommand{\AdS}{${\cal A}d\cal S$}
\newcommand{\BdS}{${\cal B}d{\cal S}$}

\newcommand{\SR}{relativity}%{${\cal SR}$}

\newcommand\btd{\raise 2pt
\hbox{$\hat\bigtriangledown$}\hskip 1.5pt}
\newcommand\bt{\raise 2pt
\hbox{$\bigtriangledown$}\hskip 1.5pt}

\newcommand{\FLT}{$\cal FLT$}

\newcommand{\E}{$\cal E$}
\newcommand{\Lo}{$\cal L$}

\omits{%\documentclass[11pt]{article}

\newcommand{\omits}[1]{}

\def\bc{\begin{center}}
\def\nno{\nonumber}
\def\ec{\end{center}}
\def\be{\begin{eqnarray}}
\def\ee{\end{eqnarray}}

\def\Dl{\Delta}
\def\La{\Lambda}

\def\dl{\delta}
\def\eps{\epsilon}

\def\si{\sigma}

\def\d#1#2{\frac{\displaystyle #1}{\displaystyle #2}}

\newcommand\btd{\raise 2pt
\hbox{$\hat\bigtriangledown$}\hskip 1.5pt}
\newcommand\bt{\raise 2pt
\hbox{$\bigtriangledown$}\hskip 1.5pt}

}

\def\no{\noindent}

\begin{document}
%\begin{document}

\title{\bf Snyder's Quantized Space-time and
De Sitter Special Relativity} %\vskip 3mm %[10mm]

\author{{Han-Ying Guo}$^{1,2}$}
%\email{hyguo@itp.ac.cn}
\author{{Chao-Guang Huang}$^{1}$}
%\email{huangcg@mail.ihep.ac.cn}
\author{{Yu Tian}$^{3}$}
\author{{Zhan Xu}$^{4}$}
%\email{zx-dmp@mail.tsinghua.edu.cn}
\author{{Bin Zhou}$^{5}$} %\email{zhoub@itp.ac.cn}
\omits{\affiliation{%
${}^1$ CCAST (World Laboratory), P.O. Box 8730, Beijing
   100080, China,}}

\affiliation{%
${}^1$ Institute of High Energy Physics, Chinese Academy of
Sciences, P.O. Box 918-4, Beijing
   100049, China, }

\affiliation{%
${}^2$ Institute of Theoretical Physics,
 Chinese Academy of Sciences,
 P.O.Box 2735, Beijing 100080, China,}
\affiliation{%
${}^3$ Department of Physics, Beijing Institute of Technology,
Beijing 100081, China}
\omits{\affiliation{%
${}^4$Department of Mathematics, Capital Normal University, Beijing
100037, China }}
\affiliation{%
${}^5$Physics Department, Tsinghua University, Beijing
   100084, China}
   \affiliation{%
${}^6$Department of Physics, Beijing Normal University, Beijing
100875, China }

\begin{abstract}
There is a one-to-one correspondence between Snyder's model in de
Sitter space of momenta and the \dS-invariant special relativity as
well as a minimum uncertainty-like relation. This indicates that
physics at the Planck length $\ell_P$ and the scale
$R=(3/\Lambda)^{1/2}$ should be dual to each other and there is
in-between gravity of local \dS-invariance characterized by a
dimensionless coupling constant $g=\ell_P/R\sim 10^{-61}$.

\bigskip

\no
\end{abstract}


\maketitle

%\tableofcontents

%\bigskip
%\newpage
\section{Introduction}

Long time ago, Snyder \cite{Snyder} proposed a quantized space-time
model in a projective geometry approach to the de Sitter (\dS)-space
of momenta with a scale $a$ near {or at} the Planck length. The
energy and momentum of a particle were identified with the
inhomogeneous projective coordinates. Then, the spacetime
coordinates  became operators $\hat x^\mu$ given by 4-`translation'
generators of \dS-algebra, being noncommutative.

Recently, \omits{in order to explain the Greisen-Zatsepin-Kuz'min
effects \cite{GZK} }the `doubly spacial relativity' or the `deformed
special relativity' (DSR) has been proposed \cite{DSR}. There is
also a large scale $\kappa$ near the Planck energy scale, related
to $a$ in \cite{Snyder}. Since some DSR models can be realized by
the identification of 4-momentum with certain coordinates on a \dS-
or \AdS-space of momenta \cite{DSRdS}, Snyder's model may be viewed
as the first of them.

The projective geometry approach is basically equivalent to the
Beltrami model \cite{beltrami} of \dS-space (\BdS). Importantly, the
Beltrami coordinates of a \dS-hyperboloid, or inhomogeneous
projective ones, {without the antipodal identification,} play a
similar role of the Minkowski coordinates in a Minkowski-space. In
these coordinates, particles and light signals move along the
timelike or null geodesics {being straight world-lines} with {\it
constant} coordinate velocities {in each patch}, respectively. Among
these systems, the properties are invariant under the fractionally
linear transformations with common denominators (\FLT s) of
\dS-group. These motions and the systems could be regarded as of inertia
without gravity.  Then, there should be the
 principle of relativity in \dS/\AdS-spacetime, respectively. Lu \cite{Lu} emphasized the issue and began to study
the special relativity in \dS/\AdS-space, with  his collaborators
\cite{LZG}. Promoted by recent observations on dark universe
\cite{darkU}, further studies have been made \cite{BdS}-\cite{T}.

In fact, in Einstein's special \SR\ the assumptions are
made\cite{1905} that rest rigid ruler is Euclidean and that time
flows itself is homogeneous. However, these are not supported by the
asymptotic behavior of  our universe \cite{darkU}. Just as weakening
the fifth axiom leads to non-Euclidean geometry, giving up {the}
assumptions leads to two kinds of the \dS/\AdS-invariant special
relativity in \dS/\AdS-spacetime, which are on an almost equal
footing with Einstein's \cite{Lu}, \cite{BdS}-\cite{c3}.

It is important that from two fundamental constants, the Planck
length $\ell_P:=(G\hbar c^{-3})^{1/2}$ and the \dS-radius
$R=(3/\La)^{1/2}$, it follows a dimensionless constant
\be \label{g}%
g:= \sqrt 3  { \ell}_P R^{-1} \quad \mbox{or} \quad g^2
= \d {G\hbar
 \Lambda} {{3}c^{3}}\sim 10^{-122}.
\ee %
As Newton constant $G$ is present in (\ref{g}), $g$ should
describe gravity. A simple gauge-like model for the \dS-gravity showed this
feature \cite{dSG, dSG2, QG}.

In this letter, we show that there is an interesting and important
one-to-one correspondence between \dS-invariant special relativity
and Snyder's model. In addition, there is also a minimum
uncertainty-like relation between them. These indicates that the
physics at the Planck scale and the scale $R$ should be dual to each
other and there is in-between the local \dS-invariant gravity
characterized by the dimensionless coupling constant $g$.

%%%%%%%%%%%%%%%%%%%%%%%%%%%%%%%%%%%%%%%%%%%%%%%%%
%
%              Beltrami model
%
%%%%%%%%%%%%%%%%%%%%%%%%%%%%%%%%%%%%%%%%%%%%%%%

\section{The Beltrami model}

The 4-d Riemann sphere ${\cal S}^4$
can be embedded in a 5-d Euclid space \E$^5$%
\be\label{4s}%
{\cal S}^4:~~\delta_{AB}\xi^A \xi^B&=&\ell^2>0, \quad A, B=0, \cdots, 4,\\\label{5ds}%
ds_E^2&=&\delta_{AB}d\xi^A d\xi^B=d\xi^t {\cal
I}d\xi. %
\ee%
where superscript $t$ represents transpose. They are invariant under
rotations of $SO(5)$:
\be\label{so5}%
 \xi~ \rightarrow ~\xi'=S \xi, \quad S^t {\cal I} S={\cal
I}, ~~\forall ~S ~\in ~{SO(5)}.
\ee%

A Beltrami model ${\cal B}$ of ${\cal S}^4$ is the intrinsic
geometry of ${\cal S}^4$ with Beltrami coordinate atlas. In a patch, say,
\be\label{Bcrd}%
x^\mu:=\ell{\xi^\mu}/{\xi^4}, \quad \xi^4\neq 0  ,\quad
 \mu=0,\cdots, 3,%
\ee%
with%
\be\label{sigma}%
\sigma_E(x):=\sigma_E(x,x)=1+\ell^{-2}\delta_{\mu \nu}x^\mu
x^\nu>0,\qquad\\\label{4Bds} ds_E^2=[\delta_{\mu
\nu}\sigma_E^{-1}(x)-\ell^{-2}\sigma_E^{-2}(x)\delta_{\mu
\si}x^\si\delta_{\nu \rho}x^\rho ] dx^\mu dx^\nu,
\ee%
it is invariant under  \FLT s of $SO(5)$ with a transitive form
sending
the point $A(a^\mu)$ to the origin $O(o^\mu=0)$,%
\be\nno
x^\mu \rightarrow \tilde{x}^\mu
&=&\pm\sigma_E^{1/2}(a)\sigma_E^{-1}(a,x)(x^\nu
-a^\nu)N_\nu^\mu,\\\label{FLT}
N_\nu^\mu &=&O_\nu ^\mu -{ \ell^{-2}}%
\delta_{\nu\si}a^\si a^\rho
[\sigma_E(a)+\sigma_E^{1/2}(a)]^{-1}O_\rho^\mu,\\\nno
O&:=&(O_\nu^\mu)\in SO(4).%
\ee %
There is an invariant for two points $A(a^\mu)$ and $B(b^\nu)$
\be\label{AB} %
{\Delta}_{E,\ell}^2(a, b) = \ell^2
[1-\sigma_E^{-1}(a)\sigma_E^{-1}(b)\sigma_E^2(a,b)].% 
\ee %
The proper length between $A$ and $B$ is the integral of ${\rm d}s_E^{}$
over the geodesic segment $\overline{AB}$:
\be \label{ABL}%
L(a,b)&=& \ell \arcsin (|\Delta_E(a,b)|/l).%
\ee %
The geodesics in ${\cal B}$ are  straight-lines and equivalent to%
\be\label{q}%
\frac{dq^\mu}{ds}=0, \quad {q^\mu}:=\sigma_E^{-1}(x)\frac{dx^\mu}{ds},%
\ee%
from which it follows the constant ratios%
\be\label{um}%
\frac{q^i}{q^0}=\frac{dx^i}{dx^0}=consts,~~i=1,2,3.%
\ee%
And they can be integrated out:
\be\label{sl}%
x^i(s)=\alpha^i x^0 +\beta^i, \quad \alpha^i,\beta^i=consts.%
\ee%

In view of gnomonic projection, the great circles on (\ref{4s}) are
\omits{globally} mapped to the straight-lines,  the geodesics
(\ref{sl}) in ${\cal B}$, and vice versa. It is also the case for
Lobachevski space \Lo$^4$ as the original model \cite{beltrami} is
just for the Lobachevski plane.

%%%%%%%%%%%%%%%%%%%%%%%%%%%%%%%%%%%%%%%%%%%%%%
%
%            dS-SR
%
%%%%%%%%%%%%%%%%%%%%%%%%%%%%%%%%%%%%%%%%%%%%%%

\section{\dS-invariant special relativity via an inverse Wick rotation}
From an inverse Wick rotation of Riemann sphere
${\cal S}^4$ with $\ell=R$ \cite{IWR}, it follows%
\be \label{qiu2}
 { H}_{R}:& &\eta_{AB}\xi ^{A}\xi ^{B}\omits{-\sum
_{\alpha=1}^{3}\xi ^{\alpha}\xi ^{\alpha}-\theta\xi ^{4}\xi
^{4}}= \xi^t {\cal J}\xi= - R^2 < 0, \\
&& \ ds^2=\eta_{A B}d\xi^A d\xi^B = d\xi^t{\cal J}d\xi, %
\ee%
with ${\cal J}=(\eta_{A B})={\rm diag}(1,-1,-1,-1,-1 )$ and the
projective boundary $\partial_P H_R: \xi^t{\cal J}\xi=0$. Under %
\be%
 \xi~ \rightarrow ~\xi'=S \xi, ~~ S^t {\cal J} S={\cal
J},~~ \forall ~S ~\in ~{SO(1,4)},
\ee%
they are invariant. Great circles in ${\cal S}^4$ now become {a kind
of} uniform `great-circular'
motions with a conserved 5-d angular momentum on $H_R$%
\be\label{angular5a} %
\frac{d{\cal L}^{AB}}{ds}=0,\quad {\cal
L}^{AB}:=m_{R}^{}\left(\xi^A\frac{d\xi^B}{ds}-\xi^B\frac{d\xi^A}{ds}\right),
\ee %
with an Einstein-like formula for mass $m_{R}^{}$
\begin{eqnarray}\label{emla}%
{ -\frac{1}{2R^2}{\cal L}^{AB}{\cal L}_{AB}=m_{R}^2,}\quad% [0mm]\nno
{\cal L}_{AB}^{}=\eta_{AC}^{}\eta_{BD}^{}{\cal L}^{CD}.
\end{eqnarray}
Further, a `simultaneous' 3-hypersurface %
\be\label{s3}%
\delta_{ab}\xi^a\xi^b=R^2+(\xi^0)^2,~~ a, b=1,\cdots, 4,
\ee %
is an expanding $S^3$.

The 5-d angular momentum operators, proportional to the generators
of the \dS-algebra $so(1,4)$ or the Killing vector fields acting on the \dS-hyperboloid, read ($\hbar=1$)
\be\label{Generator}%
 {{\hat {\cal L}}}_{AB} =\frac 1 i \left(\xi_A
\frac{\partial}{\partial\xi ^B} - \xi_B
\frac{\partial}{\partial\xi ^A}\right),~~\xi_A^{}=\eta_{AB}\xi^B.%
\ee%
{And they are globally defined on the \dS-hyperboloid.}

 Via
the inverse Wick rotation, the Beltrami model of Riemann-sphere
becomes the \BdS-space covered with Beltrami coordinate {atlas}
patch by
patch.  The condition and Beltrami metric with $\eta_{\mu \nu}%
={\rm diag} (1,-1,-1,-1)$ in each patch%
\be\label{domain}%
\sigma(x)=\sigma(x,x):=1-R^{-2} \eta_{\mu \nu}x^\mu x^\nu
>0,\qquad\\\label{metric} %
 ds^2=[\eta_{\mu\nu}\sigma^{-1}(x)+ R^{-2} \eta_{\mu\si}\eta_{\nu\rho}x^\si x^\rho
\sigma^{-2}(x)]dx^\mu dx^\nu, %
 \ee%
are invariant under $\cal FLT$s of $SO(1,4)$
\be\label{G}%
 x^\mu\rightarrow \tilde{x}^\mu &=&\pm
\sigma^{1/2}(a)\sigma^{-1}(a,x)(x^\nu-a^\nu)D_\nu^\mu,\\\nno
D_\nu^\mu&=&L_\nu^\mu - { R^{-2}}%
\eta_{\nu \si}a^\si a^\rho
[\sigma(a)+\sigma^{1/2}(a)]^{-1}L_\rho^\mu,\\\nno
L&:=&\{L_\nu^\mu\} \in SO(1,3). %
\ee
In such a \BdS-space, the generators of $\cal FLT$s, or the Killing
vectors, read
\be\label{generator}%
  {\hat q}_\mu =(\delta_\mu^\nu+R^{-2}x_\mu x^\nu) \partial_\nu,~~
  x_\mu:=\eta_{\mu\nu}x^\nu,\qquad\\\nno
  {\hat L}_{\mu\nu} = x_\mu {\hat q}_\nu - x_\nu {\hat q}_\mu
  = x_\mu \partial_\nu - x_\nu \partial_\mu \in so(1,3), %  \\\nno
\ee%
and form an $so(1,4)$ algebra %
\be\nno%
  [ \hat{q}_\mu, \hat{q}_\nu ] = - R^{-2} \hat{L}_{\mu\nu},~~  %
  {[} \hat{L}_{\mu\nu},\hat{q}_\si {]} =
    \eta_{\nu\si} \hat{q}_\mu - \eta_{\mu\si} \hat{q}_\nu,
\label{so14}\\
  {[} \hat{L}_{\mu\nu},\hat{L}_{\si\rho} {]} =
    \eta_{\nu\si} \hat{L}_{\mu\rho}
  - \eta_{\nu\rho} \hat{L}_{\mu\si}
  + \eta_{\mu\rho} \hat{L}_{\nu\si}
  - \eta_{\mu\si} \hat{L}_{\nu\rho}.
\end{eqnarray}

There are inertial motions with a set of  conserved observable
along geodesics %
\begin{eqnarray}\label{angular4}
\frac{dp^\mu}{ds}&=&0 \quad \mbox{with} \quad
p^\mu=\sigma^{-1}(x)m_{R}\frac{d x^\mu}{ds},
\\ \frac{dL^{\mu\nu}}{ds}&=&0 \quad \mbox{with} \quad  L^{\mu\nu}=x^\mu p^\nu-x^\nu p^\mu
\end{eqnarray}
or equivalent to %, as the counterparts of (\ref{eleq}),
\be\label{eleq1}%
m_R\frac{ d^2 x^i}{dt^2}=0, \qquad t=x^0/c.%
\ee
The pseudo 4-momentum $p^\mu$ and pseudo 4-angular-momentum
$L^{\mu\nu}$ constitute a conserved 5-d angular momentum
(\ref{angular5a}).  Obviously, Eq.(\ref{angular4}) is the
counterpart of (\ref{q}) and Eq.(\ref{emla}) becomes
\begin{eqnarray}\label{eml}%
E^2=m_{R}^2c^4+{p}^2c^2 + \d {c^2} {R^2} { l}^2 - \d {c^4}{R^2}{
k}^2,%
\end{eqnarray}
which is a generalized Einstein's formula with energy $E=cp^0$,
momentum $p^i$, $p_i=\dl_{ij} p^j$, boosts $k^i=L^{0i}$,
$k_i=\dl_{ij} k^j$ and 3-angular momentum $l^i=\frac 1 2\eps^{i}_{\
jk}L^{jk}$, $l_i=\delta_{ij} l^j$.  {It can be proved that they are
Noether's charges with respect to the Killing vectors
(\ref{generator}).

 It should be emphasized that since the generators in %Eq.
(\ref{Generator}) are globally defined on the \dS-hyperboloid, they
should also be globally defined in the Beltrami atlas patch by
patch. Thus there is a set of globally defined ten Killing vectors
in the Beltrami atlas and correspondingly, there is a set of ten
Noether's charges forming a 5-d angular momentum ${\cal L}^{AB}$
in (\ref{angular5a}) {\it globally} in the Beltrami atlas, though the
physical meaning of each Noether's charge depends on the Beltrami coordinate patch used.}

The interval and thus light-cone can be well
defined by the counterparts of (\ref{AB}) and (\ref{ABL}).

Thus, \dS-invariant special relativity can be set up on the
relativity principle \cite{Lu, LZG} and the  universal constant
postulate for the speed of light $c$ and radius $R$ \cite{BdS}.

%%%%%%%%%%%%%%%%%%%%%%%%%%%%%%%%%%%%%%%%%%%%%%%%%
%
%              QST model
%
%%%%%%%%%%%%%%%%%%%%%%%%%%%%%%%%%%%%%%%%%%%%%%%

\section{Snyder's model and DSR}

Snyder considered the homogenous quadratic form
%like (\ref{qiu2}),
\be\label{snyder}%
 -\eta^2:=\eta^{AB}\eta_A\eta_B<0. \ee
It may be regarded as a hyperboloid in 5-d space of momenta with
the line-element,
\be \label{dsp}%
ds_p^2 =\eta^{AB}d\eta_A d\eta_B,
\ee
and is identical to the inverse Wick rotation of
(\ref{4s}) after identifying $\eta_A$ with $\rho \xi_A$ with a
common factor $\rho\neq 0$  in relativistic units $[\rho]=L^{-2}$.
Thus, a Beltrami model of \dS-space of momenta may also
be set up on a space of
momenta. In fact,
Snyder defines the energy-momentum with help of a constant
$a$, %of length dimension
which may be taken as the Planck length,
\be\label{BdSp}%
p_0^{} =a^{-1} {\eta_0}/{\eta_4}=a^{-1} {\xi^0}/{\xi^4},\quad%&&
p_i^{} =a^{-1} {\eta_i}/{\eta_4}=a^{-1} {\xi^i}/{\xi^4}. \nno%
\ee%
Quantum mechanically, in this `momentum picture' the
operators for the space-time-coordinates
 $ \hat x^i, \hat t$ should be given by:%
\be\label{xt}%
\hat x^i &:=&i[\frac{\partial}{\partial p_i^{}}+ a^2 p^i
p_\nu^{}\frac{\partial}{\partial p_\nu}^{}],~ \\\nno%
\hat t =\hat x^0/c&:=&\frac i c[\frac{\partial}{\partial p_0^{}}+a^2
p^0
p_\nu^{}\frac{\partial}{\partial p_\nu^{}}], \quad p^\mu=\eta^{\mu\nu}p_\nu^{}%
\ee%
Together with `boost' $\hat M^i= \hat x^0p^i+\hat x^i p^0$
and `3-angular momentum' $\hat L^i=\frac 1 2 \eps ^{i}_{\ jk}\hat
x^j p^k
$, they form an $so(1,4)$ algebra %
\be\label{so14m}%
[\hat x^i, \hat x^j]=-i a^2\eps_k^{\ ij}L^k,&&%\\\nno[2mm]
[\hat x^0, \hat x^i ]=-i a^2\hat M^i, \\\nno%
[\hat L^i, \hat L^j]=\epsilon_k^{\ ij}\hat L^k,&& [\hat M^i, \hat
M^j]=\epsilon_k^{\ ij}\hat M^k;~~
etc.%
\ee%

Since $p^\mu$ as inhomogeneous (projective) coordinates or
Beltrami coordinates, one patch in the model is not enough.
And since the projective space $RP^4$ is non-orientable, to
preserve the orientation the antipodal identification should not
be taken. The operators $\hat x^\mu$ are just 4-generators of the
\dS-algebra (\ref{generator}). And $\hat L^i, \hat M^i$ are rest
6-generators $\hat L_{\mu\nu}$ in (\ref{so14}) of $so(1,3)$
algebra. Actually, the algebra (\ref{so14m}) is the same as
(\ref{so14}).

Similar to Snyder's model, a quantized space-time model on
\AdS-space of momenta can be
constructed.  Actually, some other DSR models can also be
described in other coordinates  in a \dS- or \AdS-space of momenta \cite{DSRdS}.

It is important that the correspondence of the ratio (\ref{um}) in
\dS-space of momenta may be viewed as the inverse of `group
velocity' components of some `wave-packets'.  If so, one may define
in Snyder's model a new kind of uniform motions with constant
component `group velocity'. In particular, when the correspondences
of $\beta^i$ in (\ref{sl}) vanish, the `group velocity' of a
`wave-packet' coincides its `phase velocity'.  This is similar to
the case for a light pulse propagating in vacuum  Minkowski
spacetime.

Furthermore, \dS-space of momenta also has a horizon.  Thus, one
may imitate the study of \dS-space in  general relativity to
introduce `temperature' $\tilde{T}_p$ and `entropy' $\tilde{S}_p$
for the horizon.  But the question is, do they make sense?

%%%%%%%%%%%%%%%%%%%%%%%%%%%%%%%%%%%%%%%%%%%%%%%%%
%
%              Thermodynamics
%
%%%%%%%%%%%%%%%%%%%%%%%%%%%%%%%%%%%%%%%%%%%%%%%

\section{On Thermodynamics}

 In the viewpoint of \dS-invariant special
relativity there is no gravity in \dS-space. Therefore, the
thermodynamics is not originated  from gravity.

Since there exist inertial motions and inertial observers in
\dS-invariant special relativity, one may set up inertial
reference frame.  In the viewpoint of inertial observers in an
inertial reference frame, the horizon in \dS-space is at $T=0$
without entropy.  The temperature $T=\hbar c /(2\pi Rk_B)$ and
entropy $ S=4\pi R^2c^3k_B/(G\hbar)$ in the static \dS-coordinates
or other coordinates arise from
non-inertial motions and/or non-inertial parameterization rather
than gravity \cite{T}.

Similarly, $\tilde{T}_p$ and $\tilde{S}_p$ in Snyder's
model vanish even if the horizon in the \dS-space of momenta
exists.  Thus, we may circumvent the difficulty in the
explanation of the physical meaning of $\tilde{T}_p$ and
$\tilde{S}_p$ in Snyder's model. However,  these quantities do
exist in some DSR models in \dS-space of momenta and DSR advocators
may have to face the problem of how to explain their
physical meaning.

%%%%%%%%%%%%%%%%%%%%%%%%%%%%%%%%%%%%%%%%%%%%%%%%%%
%
%              Duality
%
%%%%%%%%%%%%%%%%%%%%%%%%%%%%%%%%%%%%%%%%%%%%%%%%%%
\section{The Planck scale-$\Lambda$ duality}

It is straightforward to see that there is {an interesting and
important one-to-one correspondence} between Snyder's model and the
\dS-invariant special relativity as shown in the following
Table:\medskip
\bc
\begin{tabular}{rcl}
\hline
{\quad \dS\, special relativity} ~& &~{ Snyder's model \quad}\\
\hline
coordinate `picture'~& &~ momentum `picture' \\
\BdS-spacetime~& &~\BdS-space of momenta \\
$R \sim$ cosmic  radius ~& &~ $1/a \sim$ Planck momentum\\
constant 3-velocity ~& &~ {constant `group velocity'}\\
`quantized' momenta ~& &~ quantized space-time\\
$\hat p_\alpha, \quad \hat E\quad$ ~& &~ $\quad\hat x_\alpha,\quad \hat t$\\
$T=0$ {without} $S$ ~& &~ $ \tilde{T}_p=0$ {without} $\tilde{S}_p$\\
{No gravity}~& &~ {No gravity}\\
\hline
\end{tabular}\medskip
\ec

The one-to-one correspondence should not be considered to happen
accidentally.

In fact, there is also  a minimum uncertainty-like relation
between them and indicate why there should be the one-to-one
correspondence. We now present an argument for the relation.
Quantum mechanically, the coordinates and momenta cannot be
determined exactly at the same time if the uncertainty principle,
which reads
\be \Dl \xi^I \Dl \eta_I^{} \geq \hbar , %
 \ee
where $I=1, \cdots, 4$ and the sum over $I$ is not taken, is
valid in the embedded space\footnote{Here we simply employ
the same notation of some observable for the expectation value of
its operator over wave function in quantum mechanics.}. Limited on
the hyperboloid in embedded space, $\Dl \xi^I \leq R$. Suppose that
the momentum $\eta_I^{}$ conjugate to $\xi^I$ also takes values on a
hyperboloid. Then, $\Dl \eta_I^{}\leq \eta$ and the uncertainty
relation implies $R \eta \sim \hbar$. Here $R$ and $\eta$ are two free
parameters.  We may write it in a covariant
form
\be \label{mu}
\eta_{AB}\xi^A \xi^B\eta^{CD}\eta_C \eta_D = \hbar^2
\ee
and refer it as an uncertainty-like relation. When
the size of hyperboloid in the space of coordinates is Planck
length, namely,
\be
\eta_{AB}\xi^A \xi^B = - \ell_P^2 = -G\hbar c^{-3},
\ee
the hyperboloid in the space of momenta then has Planck scale,
\be \eta^{AB}\eta_A \eta_B =  - E_P^2/c^2 =  -\hbar
c^3/G <0, \ee
which is equivalent to the Snyder's relation (\ref{snyder}).  On the contrary, when the scale of hyperboloid in the space of momenta is %
\be
\eta^{AB}\eta_A \eta_B =- \frac  {\La\hbar^2} 3,
\ee
then we have relation (\ref{qiu2}).  Therefore, the relation
(\ref{mu})  may indicate a kind of the UV-IR
connection and the correspondence listed in the Table should reflect
some dual relation between the physics at these two scales.
Of course, the argument here should be further demonstrated. We will
provide it in detail elsewhere.

Furthermore, both Snyder's model and the \dS-invariant special
relativity deal with the motion of relativistic particles.  In
\dS-invariant special relativity, the momenta of a particle are
quantized and noncommutative, while in  Snyder's model, the
coordinates of a particle are quantized and noncommutative.  In both
of them there is no gravity.  As was mentioned at beginning,
however, the dimensionless constant $g =\ell_P/R$ in (\ref{g})
contains the gravitational constant and thus should describe some
gravity.  Therefore, we may make a
conjecture that the physics at such two scales should be dual to
each other in some `phase' space and there is in-between
the gravity characterized by $g$.

%%%%%%%%%%%%%%%%%%%%%%%%%%%%%%%%%%%%%%%%%%%%%%%%%%
%
%              Gravity
%
%%%%%%%%%%%%%%%%%%%%%%%%%%%%%%%%%%%%%%%%%%%%%%%%%%
\section{How to describe the gravity characterized by $g$?}

It is the core of the equivalence principle that the gravity should
be based on localized special \SR.  In general relativity, however,
there are only local $so(1,3)$ Lorentz frames without local
translations. One may expect that the gravity should be based on the
equivalence principle with full localized symmetry of special \SR,
similar to the gauge principle, and be governed by a gauge-like
dynamics.

Now, there are three kinds of special relativity on Minkowski, \dS\
and \AdS\ space with  Poincar\'e, \dS\ and \AdS\ group,
respectively.  Thus, there should be three kinds of gravity with
relevant localized special relativity with full local symmetry.

{These requirements have been indicated  by} a kind of simple models
of \dS/\AdS-gravity \cite{dSG, dSG2, QG}. {For the \dS-model,} the
gauge-like action with the constant $g$ of \dS-gravity in Lorentz
gauge reads \cite{dSG}
%
%\begin{widetext}
\be \nno%
S_G&=& -\d {\hbar}{4 g^2} \int d^4 x e ( {\cal
F}^{AB}_{~~~\mu\nu} {\cal F}_{AB}^{~~\,\mu\nu})\\%
& =&\int d^4x e \left (  \d {c^3} {16\pi G} (F-2\La) -\d {
\hbar}{4g^2} F^{ab}_{~~\mu\nu} F_{ab}^{~~\mu\nu}+ \frac {c^3} {32\pi
G}
T^{a}_{~\mu\nu}T_a^{~\mu\nu} \right ),%
\label{aog}
\ee
%
%\end{widetext}
where $e=det(e^a_\mu)$, ${\cal F}^{AB}_{~~\mu\nu}$ is the curvature
of a \dS-connection ${\cal B}^{AB}_{~~\mu}\in so(1,4)$, with $ {\cal
B}^{ab}_{~~\mu}=B^{ab}_{~~\mu},  {\cal
B}^{a4}_{~~\mu}=R^{-1}e^a_{~\mu}$, $F$, $F^{ab}_{~~\mu\nu}$
 and $T^{a}_{~\mu\nu}$ Cartan's scalar
curvature, curvature and torsion, respectively, on Riemann-Cartan
manifolds with metric $g_{\mu\nu}=\eta_{ab}e^a_\mu e^b_\nu$, Lorentz
frame and connection
$e^a_{~\mu}, B^{ab}_{~~\mu}\in so(1,3)$. 

{It can be shown that the \dS-space and thus the \dS-invariant
special relativity} \omits{Recent studies show that both the
\dS-space and the Schwarzschild-\dS-space} do fit this model. In
addition, the terms in the action other than the Einstein-Hilbert
term $R$, which can be picked up from the Einstein-Cartan term $F$,
should play an important role as some `dark matter' in the viewpoint
of general relativity. Thus, this model may provide a {new}
\omits{more suitable}platform for the data analysis of dark
universe.

To show whether the the Snyder's model also fits the model of
gravity, one needs to study the quantization of the model of gravity in a nonperturbative
procedure.  Undoubtedly, there is a long way to go.  Fortunately, it
has been shown that the  model of gravity is renormalizable {perturbatively}
\cite{QG}.  {Also, the} \omits{its}Euclidean {version of} action
(\ref{aog}) is $so(5)$ gauge-like {and} \omits{having}the Riemann
sphere {is its solution} as an instanton.  So, the quantum tunneling
scenario should support $\Lambda>0$. Furthermore, asymptotic freedom
may imply that the coupling constant $g$ should be very tiny and it
should link $\Lambda$ as an infrared cut-off with $\ell_P$ as an
ultraviolet cut-off providing a fixed point.

Finally, note that $g^2$ is in the same order of difference
between $\Lambda$ and the theoretical quantum `vacuum energy', the
big difference is no longer a puzzle in the viewpoint of the
\dS-invariant special relativity and local \dS-invariant gravity.
Since $\Lambda$ is a fundamental constant as $c$, $G$ and $\hbar$, a
further question should be: what are the origins of these
fundamental constants or the origin of the dimensionless constant
$g$ and is $g$ calculable?

%%%%%%%%%%%%%%%%%%%%%%%%%%%%%%%%%%%%%%%%%%%%%%%

\section{Concluding Remarks}

We have shown the one-to-one correspondence between Snyder's model
and the \dS-invariant special relativity as well as the minimum
uncertainty-like relation. Based on this correspondence and the
relation, we have made a conjecture that there should be a duality
in physics at the Planck scale and at the cosmological scale $R$ and
that there is in-between gravity characterized by a dimensionless
constant $g$.

The gravity between the two scales should be based on the
localization of the \dS-invariant special relativity with a
gauge-like dynamics. A simple model of \dS-gravity in the Lorentz
gauge may support this point of view.

\vskip 2mm

We would like to thank Professors Z. Chang, Q.K. Lu and Z. Zhao for
valuable discussions. This work is partly supported by NSFC under
90403023, 90503002, 10505004, 10547002 and 10605005 and Knowledge
Innovation Funds of CAS (KJCX3-SYW-S03).

\end{document}